\documentclass[12pt]{article}
\usepackage{times,amssymb,euscript,epsfig,latexsym,amsmath}
\usepackage[normalem]{ulem}
\usepackage{epsfig,graphicx}
\usepackage{color}
\begin{document}
\newcommand{\todo}[1]{\vspace{5 mm}\par \noindent
\framebox{\begin{minipage}[c]{0.95 \textwidth}
\tt #1 \end{minipage}}\vspace{5 mm}\par}
\newtheorem{theorem}{Theorem}[section]
\newtheorem{corollary}{Corollary}
\newtheorem{lemma}[theorem]{Lemma}
\newtheorem{proposition}{Proposition}
\newtheorem{proof}{Proof}[section]
\date{}
\title{ \Large{On the Hamiltonian structure of the intrinsic evolution of a closed vortex sheet}}
\author{Banavara N. Shashikanth\footnote{Mechanical and Aerospace Engineering
Department, MSC 3450, PO Box 30001, New Mexico State University,
Las Cruces, NM 88003, USA. E-mail:shashi@nmsu.edu}}

\maketitle

\abstract{Motivated by the work of previous authors on vortex sheets and their applications, the intrinsic inviscid evolution equations of a closed vortex sheet in a plane, separating two piecewise constant density fluids, and their Hamiltonian form are investigated. The model has potential applications to problems involving the dynamics of interfaces of two immiscible fluids. A boundary Poisson bracket, which appears to be new and related to the KdV bracket, is obtained containing the curve-tangential derivative $\partial / \partial s$. Lagrangian invariants of the sheet motion by its self-induced velocity--the Cauchy principal value of the Biot-Savart integral--are also derived.}

\tableofcontents

\newpage

\section{Introduction.} A vortex sheet is a mathematical model of a contact discontinuity in an inviscid fluid  flow across which there is a jump in the tangential velocity. Geometrically, it is a co-dimension-1 surface. Sheets could be open or closed; this paper studies a closed vortex sheet in the plane. The jump in the tangential velocity leads to a Dirac delta distribution $\delta_D$ of vorticity supported on the surface. In the plane, the vorticity 2-form associated with a vortex sheet can be written as
\begin{align}
\omega(r,t)&=\gamma(s,t)\delta_D(r-r(s))dx \wedge dy,  \quad r \in \mathbb{R}^2, \: r(s) \in C, \label{eq:vort2form}
\end{align}
where $C$ is the curve representing the sheet, $s$  is the curve parameter and $\gamma$ is the time-varying vortex sheet strength distribution. By the Biot-Savart integral for incompressible  velocity fields of vorticity distributions, the motion of the sheet thus generates unsteady irrotational flows in the fluid domain.

Vortex sheets have found applications in a wide variety of fluid flows. In aerospace engineering, for example, they have been used to model both simple and complex flows of Newtonian fluids (ex: air and water) over lifting bodies. A `bound vortex'--a stationary vortex sheet wrapped around the surface of a lift-generating aircraft wing in a steady stream--has been traditionally used to model boundary layers and enforce the Kutta trailing edge condition in an inviscid framework. The vortex panel method, a simple computational algorithm based on the bound vortex notion, is more or less a standard topic in aerodynamic courses \cite{An1991}.  Dynamically evolving vortex sheets have been used to model the more complex phenomenon of vortex shedding from both fixed and flapping airfoils \cite{ShEl2007}. In particular, the problem of roll-up of open vortex sheets  has been a well-studied problem from both theoretical and computational perspectives \cite{BiFi1959, Mo1974, Mo1979, Sa1992}.

      Other important areas of application are the dynamics of two-fluid non-compact interfaces (open vortex sheets) \cite{Sa1992, BeBr1997, MeBaOr1982, SuSuBaFr1981, SuSu1985}, and bubble dynamics (closed vortex sheets) \cite{baker, kao}. The latter topic is a subject in itself, with or without the use of vortex sheets, and there are innumerable papers on the topic. Solutions of the vortex sheet evolution equation have also been investigated for existence and regularity properties using functional analytic tools \cite{SuSuBaFr1981, CaOr1986, CaOr1989, BaLa2010}.

  The traditional way of evolving a vortex sheet is through the Birkhoff-Rott equation which is an integro-differential equation for the curve position \cite{Sa1992}. However, it does not provide an explicit evolution equation for $\gamma(s,t)$, an issue circumvented by expressing the contour integral in B-R using a curve differential parameter that is a Lagrangian invariant. Sulem {\it et al} \cite{SuSuBaFr1981} were possibly the first to come up with a system of 1st order PDE for the simultaneous evolution of both the sheet position and $\gamma$, for the case of 2D and 3D homogeneous fluids. The PDEs of course still have to be tagged with the Biot-Savart integral. Baker, Meiron and Orszag \cite{bmo} developed evolution equations for a vortex sheet separating two fluids with different densities. Subsequently, Sulem and Sulem \cite{SuSu1985} also considered such a system, and wrote their equations using a density-weighted sheet strength distribution variable. Hamiltonian structure of the PDEs was not addressed in these papers. More recently, Izosimov and Khesin have derived vortex sheet evolution equations and Hamiltonian structure working in the more abstract setting of diffeomorphism groupoids \cite{IzKh2018}, making connections to the geometric mechanics formulations of Arnold \cite{Ar1966}, Marsden and Weinstein \cite{MaWe1983} and Lewis, Marsden, Montgomery and Ratiu \cite{LeMaMoRa1986}. 

\begin{figure}
\centering
\vspace{-1in}
\includegraphics[width=3in]{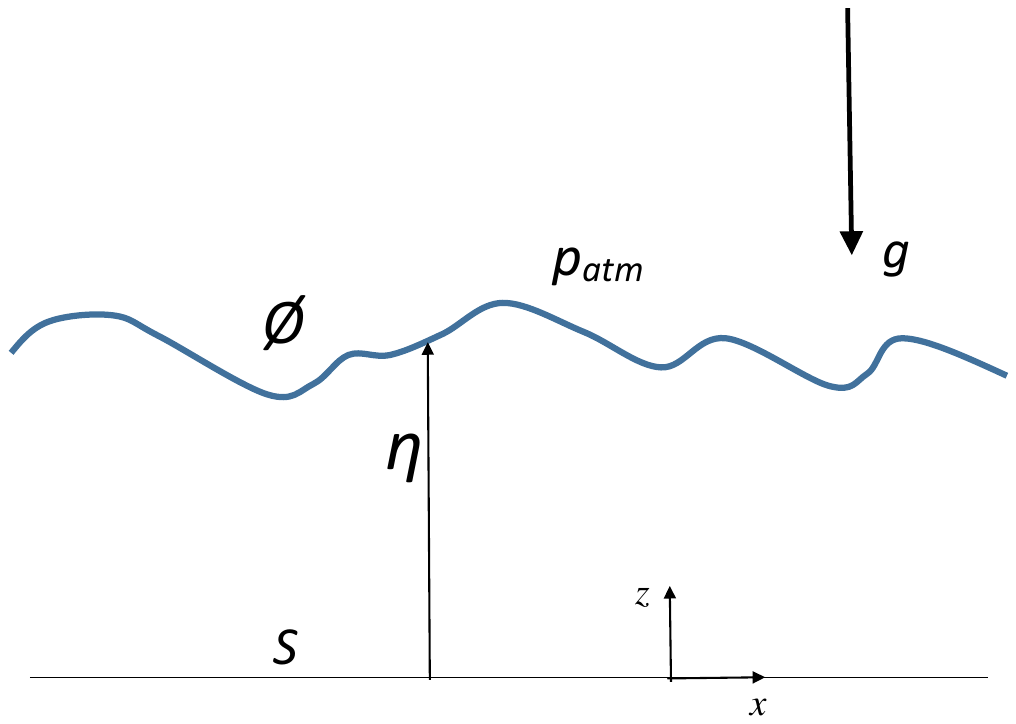}
\vspace{-1in}
\caption{Schematic sketch of the 3D free surface problem considered by Zakharov }
\label{Zakh}
\end{figure}

\paragraph{Zakharov's problem and vortex sheets.} Another time-honored formulation of boundary motion that generates unsteady irrotational flows is due to Zakharov \cite{Za1968}. In his seminal work on the Hamiltonian formalism of inviscid, irrotational, free surface water waves,  he presented a canonical Poisson bracket for the nonlinear evolution of the waves  in the variables $(\eta,\phi)$, taking into account the effects of surface tension. Figure \ref{Zakh} shows a 2D schematic of the problem, $\eta$ represents the vertical height of the waves above a horizontal datum and $\phi$ is the restriction to the free surface of $\Phi$, the velocity potential function of the irrotational flow in the domain. The coupled system of first-order nonlinear PDE is propagated in time in conjunction with a Dirichlet or/and Neumann problem which updates $\Phi$ in the domain at every time step. The motion of the free surface thus generates the entire irrotational flow in the domain. The problem can also be formulated in terms of the variables $(\vec{\Sigma},\phi)$, where $\vec{\Sigma}$  locates the instantaneous position of the parametrized sheet in $\mathbb{R}^2$ and is a more convenient choice when applying the Zakharov formulation to compact domains \cite{Sh2016}.  

Benjamin and Bridges \cite{BeBr1997} extended Zakharov's free surface water-wave problem \cite{Za1968} to a two-fluid interface problem by taking explicit account of the different densities, with the addition of a minor detail--a constant uni-directional wind. They used a density-weighted velocity potential function--the `conjugate momentum' variable-- and presented the equations of motion for the problem in canonical Hamiltonian form, using variables similar to the canonical pair used by Zakharov; see also \cite{Am2000}. Craig, Guyenne and Kalisch \cite{CrGuKa2005} consider a similar problem but starting first from the Lagrangian setting.  

%   Zakharov's problem can also be viewed as a two-fluid interface problem if one takes into account the dynamics of the air above the water, under the same set of assumptions. For an upper layer of finite depth the free surface condition gets replaced by a pressure continuity condition. 

 It is natural to expect a close link between vortex sheet and Zakharov formulations, but seemingly this has not been well explored in the literature. The main result of this paper is to  use this relation to derive a vortex sheet Poisson bracket from Zakharov's canonical brackets, which appears to be new. An intrinsic Hamiltonian formulation is obtained in which coupled first-order evolution equations for the sheet are coupled with Biot-Savart integrals, with all terms evaluated on the sheet itself. The intrinsic nature of vortex sheet formulation, not unknown to previous researchers, gives it an advantage over Zakharov's formulation since the computation of spatial domain operators in the Dirichlet/Neumann problem is avoided.  Hamiltonian formulations are important both from theoretical and computational perspectives. From a theoretical perspective, especially from the perspective of the Hamiltonian structure of singular co-dimension-1-and-2 vortex models, this structure fits  nicely alongside the well-known intrinsic Hamiltonian structure for co-dimension 2 models, namely, point vortices in $\mathbb{R}^2$ and vortex filaments/rings in $\mathbb{R}^3$ \cite{MaWe1983}. From a computational perspective, growing research in simulations of differential equations has highlighted the importance of developing numerical codes that preserve underlying fundamental structures like Poisson brackets, symplectic forms and metric structures (\cite{MaPaSh1998, Eldred2017}). These codes now go under various rubrics: `structure-preserving algorithms', `Poisson integrators' and `symplectic integrators'. Developing such integrators for the model in this paper, consistent with its Poisson bracket structure, could lead to improved simulations of vortex sheet motion. 

      In this paper, a closed vortex sheet in the plane is considered, with pressure continuity imposed across the interface in the absence of surface tension  (though this effect could be included if needed). The configuration is similar to the two-fluid interface model of \cite{BeBr1997}, except that `top fluid' and `bottom fluid' are replaced by `outside fluid' and`inside fluid', with the latter occupying a compact domain. Evolution equations are derived using a density-weighted $\gamma$-variable similar to Sulem {\it et al} \cite{SuSuBaFr1981}. The vortex sheet problem is viewed, in a novel manner, as the sum of two Zakharov problems which leads to a canonical Poisson bracket structure defined on the sheet containing the curve-tangential derivative $\partial / \partial s$ and bears resemblance to the famous KdV bracket \cite{KdV1895, GaGrKrMu1967, Mi1981}. Finally, Lagrangian invariants of the motion, following approaches similar to Kelvin's circulation theorem, are derived and discussed. An Appendix with a brief introduction to the mathematical formalities associated with infinite-dimensional Hamiltonian systems is presented for readers unfamiliar with this formalism.

    As noted by previous researchers, closed vortex sheets models have potential applications to the dynamics of gas bubbles undergoing shape deformation, provided compressibility effects can be neglected. More appropriate applications would be to configurations of a liquid in another immiscible liquid, for example, oil injected in water or an oil slick on or near the water surface. Another potential application is to the problem of a liquid drop falling in a  gaseous medium. However, the neglect of viscous effects in the model would require some justification in applications to bubbles and drops of small sizes.

\section{Hamiltonian structure for a closed vortex sheet via Zakharov's canonical bracket.} 

%\paragraph{Mathematical background.} 

%Poisson and symplectic manifolds bear an intimate relation to each other. Any symplectic  manifold is also a Poisson manifold, with the symplectic form and Poisson bracket related as: 
%\[\Omega \left(X_F,X_G \right)(p)=\left\{F,G \right\}(p) \]
%More generally, a Poisson manifold consist of symplectic submanifolds termed symplectic leaves. 

  We begin by recalling some basic facts about vortex sheets \cite{Sa1992}.  It will be assumed thoughout this paper that $s$ is the arc-length parameter. The notation $C$ is used to denote the set of points that lie on the sheet.  The velocity field in $\mathbb{R}^2$ due to the vortex sheet is given by the Biot-Savart integral,
\begin{align}
v(q)&=-\frac{1}{2 \pi} \oint_{C} \gamma (\tilde{s},t) \hat{k} \times  \frac{ r - r(\tilde{s})}{\mid r - r(\tilde{s}) \mid^2} \; d \tilde{s}, \quad q \in \mathbb{R}^2 \label{eq:bsvel}
\end{align}
where $r$ is the position vector of the point $q$. 
In particular, if $q \equiv r(s) \in C$, then the integral 
\begin{align}
v(q)&=-\frac{1}{2 \pi} \oint_{C} \gamma (\tilde{s},t) \hat{k} \times  \frac{ r(s) - r(\tilde{s})}{\mid r(s) - r(\tilde{s}) \mid^2} \; d \tilde{s}, \quad q \in C, \label{eq:cpv}
\end{align}
is finite when evaluated as an improper integral and is termed the Cauchy principal value, denoted by $CPV$. The following relations hold
\begin{align}
CPV&=\frac{v_o+v_i}{2}_{\mid_C} \\
{v_{o}}_{\mid_C} &=\frac{\gamma}{2} \hat{t}_{\mid_C} +CPV \label{eq:cpvo} \\
{v_{i}}_{\mid_C}&=-\frac{\gamma}{2} \hat{t}_{\mid_C}+CPV \label{eq:cpvi} \\
v_{o}\cdot \hat{n}_{\mid_C}&=v_{i}\cdot \hat{n}_{\mid_C}=CPV \cdot \hat{n}_{\mid_C} \label{eq:normalvel}
\end{align}
with the vortex sheet strength distribution $\gamma(s,t)$ defined as 
\begin{align}
\gamma&=(v_{o}-v_{i})\cdot \hat{t}_{\mid_C} \label{eq:gammadef}
\end{align}
The above definition makes it obvious that $\gamma(s,t)$ is independent of the choice of curve parameter, since it depends on the values of the Eulerian velocity fields at the underlying domain points. Since the alternative choice of a Lagrangian curve parameter takes into account the tangential displacement of fluid elements on $C$, this implies that the tangential displacement plays no role in the dynamics of the sheet which can therefore be treated as a geometric interface rather than a material interface, just like the free surface in Zakharov's problem.

Let the irrotational flows in the two domains be governed by velocity potentials $\Phi_{o}:D_o \rightarrow \mathbb{R}$ and $\Phi_{i}\rightarrow \mathbb{R}$, respectively, and let 
\begin{align}
\phi_{o}&:={\Phi_{o}}_{\mid_C}, \quad  \phi_{i}:={\Phi_{i}}_{\mid_C} \label{eq:velpots}
\end{align}
From the previous relations, it follows that 
\begin{align}
\nabla \Phi_{o}\cdot \hat{n}_{\mid_C}&=\nabla \Phi_{i}\cdot \hat{n}_{\mid_C} \label{eq:normalvelphi} \\
\gamma&=(\nabla \Phi_{o} - \nabla \Phi_{i})\cdot \hat{t}_{\mid_C}=\frac{\partial (\phi_{o}-\phi_{i})}{\partial s} \label{eq:link} \\
\Rightarrow \delta \gamma&=\delta \left(\frac{\partial (\phi_{o}-\phi_{i})}{\partial s} \right)=\frac{\partial (\delta (\phi_{o}-\phi_{i}))}{\partial s}, \label{eq:phigammasi}
\end{align}
where $\delta$ denotes variation. Equation (\ref{eq:link}) links the vortex sheet and Zakharov formulations. The last equation is a consequence of the commutability of variations and derivatives, a standard assumption made in variational calculus \cite{La1970}.

\subsection{The case of a uniform density field. } The case of a homogeneous fluid is first considered. For this case, assume there is non-trivial flow in both domains but the densities are the same and equal to unity (wlog):
\begin{align}
\rho_o=\rho_i =1\label{eq:assumptions1}
\end{align}

\noindent{\it Two Zakharov problems.} To construct the vortex sheet bracket, we will view the two flows generated by the vortex sheet as the disjoint union of two Zakharov flows generated by the common boundary $C$, in the outer domain $D_o$ and in the inner domain $D_i$, respectively. Dynamically, this means that the two flows influence each other only through the motion of $C$ and there is no direct domain-to-domain interaction. 

 Referring to the discussion in Appendix A, let $\mathcal{Z}_o,\mathcal{Z}_i$ denote the phase space of each set of flows. Each is an infinite-dimensional manifold, with `coordinates'  $(\vec{\Sigma},\phi_o)$ and $(\vec{\Sigma},\phi_i)$, respectively, where $\vec{\Sigma}(s,t)$ locates the instantaneous position of the parametrized sheet in $\mathbb{R}^2$,  and each $\phi_o,\phi_i$ is the restriction to $C$ of $\Phi_o, \Phi_i$, respectively, the velocity potential functions of the irrotational flows in each domain. Note also that in each domain the harmonic function $\Phi_o$ or $\Phi_i$ is uniquely determined by the pair $(\vec{\Sigma},\phi_o)$ or $(\vec{\Sigma},\phi_i)$, respectively, from the uniqueness of the associated Dirichlet problem in each domain.  This set of coordinates is slightly different from the set $(\eta,\phi)$ used by Zakharov in the free surface problem. As stated in the introduction, this is a better choice for surfaces bounding compact domains. Mathematically, though $\vec{\Sigma}$ is a vector field the curve $C$ is propagated by only normal-to-curve variations which can be viewed as real-valued functions defined on $C$ \cite{LeMaMoRa1986}, as discussed more below. Note that $C$ is assumed to be {\it positively oriented boundary for the outer domain $D_o$ and, therefore, negatively oriented for the inner domain $D_i$}.
 
  Consider functionals $F_o: \mathcal{Z}_o \rightarrow \mathbb{R}$ and $F_i:\mathcal{Z}_i \rightarrow \mathbb{R}$ on each infinite-dimensional Zakharov manifold, respectively, of the form,
\begin{align*}
F_o(z_o)&:=\oint_{C} f_o (\vec{\Sigma},\phi_o)\; ds, \quad z_o \equiv (\vec{\Sigma},\phi_o) \in \mathcal{Z}_o\quad f_o:C\rightarrow \mathbb{R} \\
F_i(z_i)&:=\oint_{C} f_i (\vec{\Sigma},\phi_i)\; ds, \quad z_i \equiv (\vec{\Sigma},\phi_i) \in \mathcal{Z}_i \quad f_i: C \rightarrow \mathbb{R}.
\end{align*}
Recalling the standard expression for kinetic energy of an irrotational flow in a domain with boundary $C$, \[\frac{1}{2} \oint_C \phi \nabla \Phi \cdot \hat{n} \; ds,\]
which is an example of either $F_o$ or $F_i$, it should be noted that $f_o, f_i$ could contain domain derivatives of $\Phi_o,\Phi_i$ evaluated on $C$. Morever, this example also shows that in general $f_o \neq f_i$ since there is a jump in $\Phi$, due to the tangential sip,  as one crosses the sheet.
  
   The combined phase space of the two flows is the product space $\mathcal{Z}:=\mathcal{Z}_o \times \mathcal{Z}_i$ with real-valued functions $F_\mathcal{Z}: \mathcal{Z} \rightarrow \mathbb{R}$. Since  $(F_\mathcal{Z})_{\mid \mathcal{Z_o}}=F_o$ and $(F_\mathcal{Z})_{\mid \mathcal{Z_i}}=F_i$, it follows from (\ref{eq:diffF}) that
\begin{align*}
dF_\mathcal{Z}(\vec{\Sigma},\phi_o,\phi_i)&=\oint_{C} \frac{\delta f_o}{\delta \Sigma} \delta \Sigma \; ds+ \oint_{C} \frac{\delta f_i}{\delta \Sigma} \delta \Sigma \; ds+\oint_{C} \frac{\delta f_o}{\delta \phi_{o}} \delta \phi_{o} \; ds  +\oint_{C} \frac{\delta f_i}{\delta \phi_{i}} \delta \phi_{i} \; ds,
\end{align*}
where 
\begin{align}
\delta \Sigma&:=\delta{\vec{ \Sigma}} \cdot \hat{n} \label{eq:delsigma}
\end{align} is a real-valued function on $C$ which can also be identified with the normal vector field $\hat{n} \delta \Sigma $ based on $C$. The functional derivatives $\delta F/\delta \Sigma, \delta F/\delta \phi_o, \delta F/\delta \phi_i$ can all be identified with real-valued functions on $C$ \cite{LeMaMoRa1986, Sh2016}. 

    Moreover, referring to the discusion following equation (\ref{eq:gammadef}), the variations $\delta{\vec{\Sigma}}$ are {\it only} in the normal direction, i.e. 
\begin{align}
\delta{\vec{\Sigma}} \cdot \hat{t}&=0 \Rightarrow \delta{\vec{\Sigma}}=\delta \Sigma \hat{n} \label{eq:delsigmatan}
\end{align} (note that equation (\ref{eq:delsigma}) by itself does not imply this). Following a variation $\hat{n} \delta \Sigma$, the new curve will generally require a re-parametrization. In other words, a point on the old curve, after being displaced by this amount, will be associated with a different parameter value on the new curve. 

%A curve-tangential variation of a point $p(s) \in C$ changes only the value of the geometric parameter $s$ and has no physical signifcance. A $\delta \vec{\Sigma}$ that is tangent at all points of $C$ leaves the sheet unaltered and causes no change in the flows. It corresponds to a zero variation of the corresponding point in the infinite-dimensional manifold. This feature may contrasted with the choice of a Lagrangian parameter where a tangential variation represents the sliding of a fluid element along the sheet. 

%For a variation $\hat{n} \delta \Sigma$ of $O(\epsilon)$ the difference in parameter values will be of the same order.

To proceed from $\mathcal{Z}$ to the infinite-dimensional space of vortex sheets, denoted by $\mathcal{V}$, we perform the change of coordinates defined by relation (\ref{eq:link}), thus equipping $\mathcal{V}$  with `coordinates' $(\vec{\Sigma},\gamma)$. Let $F:\mathcal{V} \rightarrow \mathbb{R}$ denote real-valued functions of the form
\begin{align}
F(v)&:=\oint_{C} f (\vec{\Sigma},\gamma)\; ds, \quad v \equiv (\vec{\Sigma},\gamma) \in \mathcal{V} \quad f:C \rightarrow \mathbb{R}, 
\end{align}
Since this is a change of coordinates of the same mathematical space, it follows from basic calculus (see also \cite{LeMaMoRa1986}) that 
\[dF(\vec{\Sigma},\gamma)=
dF_\mathcal{Z}(\vec{\Sigma},\phi_o,\phi_i)\]
 i.e.
\begin{align}
&\oint_{C} \frac{\delta f}{\delta \Sigma} \delta \Sigma \; ds+\oint_{C} \frac{\delta f}{\delta \gamma} \delta \gamma \; ds  \nonumber \\
&\hspace{1in} =\oint_{C} \frac{\delta f_o}{\delta \Sigma} \delta \Sigma \; ds+ \oint_{C} \frac{\delta f_i}{\delta \Sigma} \delta \Sigma \; ds+\oint_{C} \frac{\delta f_o}{\delta \phi_{o}} \delta \phi_{o} \; ds  +\oint_{C} \frac{\delta f_i}{\delta \phi_{i}} \delta \phi_{i} \; ds \label{eq:totalsi}
\end{align}

 The following identity, a simple consequence of Stokes theorem on a boundaryless manifold,  will be made us of at various points hereafter. For any smooth functions $f,g:C \rightarrow \mathbb{R}$,
\begin{align*}
\oint_C \frac{\partial f}{\partial s}g\;ds&=-\oint_C \frac{\partial g}{\partial s}f\;ds
\end{align*}
The total variation equation now becomes
\begin{align*}
&\oint_C  \frac{\delta f}{\delta \Sigma} \delta \Sigma \; ds  +\oint_{C} \frac{\delta f}{\delta \gamma}\frac{\partial}{\partial s}\left( \delta \phi_{o}-  \delta \phi_{i}\right)\; ds \\
& \hspace{0.5in}=\oint_{C} \frac{\delta f_o}{\delta \Sigma} \delta \Sigma \; ds+ \oint_{C} \frac{\delta f_i}{\delta \Sigma} \delta \Sigma \; ds+\oint_{C} \frac{\delta f_o}{\delta \phi_{o}} \delta \phi_{o} \; ds  +\oint_{C} \frac{\delta f_i}{\delta \phi_{i}} \delta \phi_{i} \; ds \\ \\
&\Rightarrow \oint_C  \frac{\delta f}{\delta \Sigma} \delta \Sigma \; ds  -\oint_{C}\frac{\partial}{\partial s} \left( \frac{\delta f}{\delta \gamma}\right) \delta \phi_{o}  \; ds+\oint_{C}\frac{\partial}{\partial s} \left( \frac{\delta f}{\delta \gamma}\right) \delta \phi_{i}  \; ds\\
& \hspace{0.5in}=\oint_{C} \frac{\delta f_o}{\delta \Sigma} \delta \Sigma \; ds+ \oint_{C} \frac{\delta f_i}{\delta \Sigma} \delta \Sigma \; ds+\oint_{C} \frac{\delta f_o}{\delta \phi_{o}} \delta \phi_{o} \; ds  +\oint_{C} \frac{\delta f_i}{\delta \phi_{i}} \delta \phi_{i} \; ds 
\end{align*}
where $\delta f_o/\delta \Sigma, \delta f_i/\delta \Sigma$ are the functional derivatives in Zakharov's model.

This leads to the following relations between the functional derivatives:
\begin{align}
\frac{\delta f_o}{\delta \Sigma}+\frac{\delta f_i}{\delta \Sigma}&=\frac{\delta f}{\delta \Sigma}   \label{eq:sigmafd} \\
\frac{\delta f_o}{\delta \phi_{o}} &=-\frac{\partial}{\partial s} \left( \frac{\delta f}{\delta \gamma}\right) \label{eq:phigammafdsi1} \\
\frac{\delta f_i}{\delta \phi_{i}} &=\frac{\partial}{\partial s} \left( \frac{\delta f}{\delta \gamma}\right), \label{eq:phigammafdsi2}
\end{align}
From the above equations, we get 
\begin{align}
\frac{\delta f_o}{\delta \phi_{o}} +
\frac{\delta f_i}{\delta \phi_{i}}&=0 \label{eq:rel}
\end{align}

\paragraph{Vortex sheet bracket.} To obtain the vortex sheet bracket, we start with the following Poisson bracket on the product space $\mathcal{Z}=\mathcal{Z}_1\times \mathcal{Z}_2$ which is the sum of the Zakharov brackets on $\mathcal{Z}_1$ and $\mathcal{Z}_2$, respectively: 
\begin{align*}
\left\{F,G \right\}_Z&:=\oint_{C}\left( \frac{\delta f_o}{\delta \Sigma } \frac{\delta g_o}{\delta \phi_{o}} -\frac{\delta f_o}{\delta \phi_{o}} \frac{\delta g_o}{\delta \Sigma }\right)  \, ds -\oint_{C}  \left(\frac{\delta f_i}{\delta \Sigma } \frac{\delta g_i}{\delta \phi_{i}} - \frac{\delta f_i}{\delta \phi_{i}} \frac{\delta g_i}{\delta \Sigma } \right)  \, ds 
\end{align*}
Using the various relations  obtained above between the functional derivatives, we immediately get 
\begin{align*}
\left\{F,G \right\}_Z&=\oint_{C}\left( \left(\frac{\delta f_o}{\delta \Sigma } +\frac{\delta f_i}{\delta \Sigma }\right)\frac{\delta g_o}{\delta \phi_{o}} -\frac{\delta f_o}{\delta \phi_{o}} \left(\frac{\delta g_o}{\delta \Sigma }+\frac{\delta g_i}{\delta \Sigma }  \right)\right)  \, ds, \\
&=\oint_{C} \left[\left( \frac{\delta f}{\delta \Sigma}\right) \left(-\frac{\partial}{\partial s} \left( \frac{\delta g}{\delta \gamma}\right) \right) \right. \\
&\hspace{1in}\left. - \left( \frac{\delta g}{\delta \Sigma} \right) \left(-\frac{\partial}{\partial s} \left( \frac{\delta f}{\delta \gamma}\right) \right)\right]\, ds \\
&=\oint_{C} \left[-\frac{\delta f}{\delta \Sigma} \frac{\partial}{\partial s} \left( \frac{\delta g}{\delta \gamma}\right)+  \frac{\delta g}{\delta \Sigma} \frac{\partial}{\partial s} \left( \frac{\delta f}{\delta \gamma}\right)\right]\, ds  \\
\end{align*}
which leads to the vortex sheet Poisson bracket
\begin{align}
\left\{F,G\right\}_V(\vec{\Sigma},\gamma)&=\oint_{C} \left[-\frac{\delta f}{\delta \Sigma} \frac{\partial}{\partial s} \left( \frac{\delta g}{\delta \gamma}\right)+  \frac{\delta g}{\delta \Sigma} \frac{\partial}{\partial s} \left( \frac{\delta f}{\delta \gamma}\right)\right]\, ds \label{eq:pbvs1}
\end{align}
\paragraph{Hamiltonian function and functional derivatives.}
The Hamiltonian function is the total kinetic energy of the flow, which is finite provided one makes the assumption that 
\begin{align}
\oint_{C_R} v_o \cdot \hat{t} \; ds& \rightarrow 0, \quad R \rightarrow \infty, \label{eq:farfield}
\end{align}
where $C_R$ is circle of radius $R$ enclosing the sheet. Applying Stokes' theorem in the irrotaional flow domains $D_o$ and $D_i$ gives,
\begin{align*}
\oint_{C_R}v_o \cdot \hat{t} \; ds&=\oint_{C}v_o \cdot \hat{t} \; ds, \quad \oint_{C}v_i \cdot \hat{t}\;ds=0, 
\end{align*}
respectively. It is easily seen then that (\ref{eq:farfield}) is equivalent to the assumption that 
\begin{align*}
\Gamma&:=\oint_C\gamma \;ds=0.
\end{align*}
If $\Gamma \neq 0$, then the Hamiltonian is only the finite part of the total kinetic energy obtained by subtracting the term $\psi \Gamma/2$, where $\psi \rightarrow \Gamma \log R/4 \pi$, as $R \rightarrow \infty$, the limit being the streamfunction due to a single point vortex of strength $\Gamma$ at the origin. 

 Due to the fact that flows are irrotational in the domains, the Hamiltonian can be expressed as the sum of two contour integrals 
\begin{align}
H(\vec{\Sigma},\gamma)&:=-\frac{1}{2} \oint_{C} (\psi_o v_o- \psi_i v_i) \cdot \hat{t} \; ds, \label{eq:hamA}
\end{align}
where $\psi_o,\psi_i$ are the streamfunctions in the respective domains satisfying: 
\begin{gather} \label{eq:streamvel}
\begin{aligned}
\nabla \psi_o \cdot \hat{n}&=-v_o \cdot \hat{t}, \quad \nabla \psi_i \cdot \hat{n}=-v_i \cdot \hat{t} \\
\nabla \psi_o \cdot \hat{t}&=v_o \cdot \hat{n}, \quad \nabla \psi_i \cdot \hat{t}=v_i \cdot \hat{n}
\end{aligned}
\end{gather}
Note from the second line that these streamfunctions are continuous across $C$, due to (\ref{eq:normalvel}), i.e $\psi_i \mid_C=\psi_o \mid_C=\psi$. This allows the two contour integrals to be combined as one and makes more transparent the dependency of $H$ on $\gamma$.
\begin{align}
H(\vec{\Sigma},\gamma)&:=\oint_C h \; ds=-\frac{1}{2} \oint_{C} \psi (v_o-v_i) \cdot \hat{t} \; ds=-\frac{1}{2} \oint_{C} \psi  \gamma \; ds,  \label{eq:hamB}
\end{align}
with $\psi$ viewed as a functional of $\gamma$, as implied by relations (\ref{eq:streamvel}).  Variations in $\gamma$, keeping $C$ constant, therefore induce variations $\delta^\gamma \psi$ in $\psi$, which are again harmonic functions in $D_o$ or $D_i$ as the case may be. To obtain the functional derivative $\delta h /\delta \gamma$, substitute (\ref{eq:streamvel}) in (\ref{eq:hamA}),
\begin{align}
H(\vec{\Sigma},\gamma)&:=\oint_C h \; ds=\frac{1}{2} \oint_{C}\left( \psi_o \nabla \psi_o \cdot \hat{n}-\psi_i \nabla \psi_i \cdot \hat{n} \right) \; ds
\end{align}
Making the usual assumption of variations commuting with spatial operators and using the following integral relation that follows from Stokes' theorem: for $f,g$ harmonic in $D_o$ or $D_i$:
\begin{align*}
\int_{C} f \nabla g \cdot \hat{n} \; ds&=\int_{D_o} \nabla f \cdot \nabla g \; dA=\int_{C} g \nabla f \cdot \hat{n} \; ds, \quad f,g: D_o \rightarrow \mathbb{R}, \\
\int_{C} f \nabla g \cdot \hat{n} \; ds&=-\int_{D_i} \nabla f \cdot \nabla g \; dA=\int_{C} g \nabla f \cdot \hat{n} \; ds, \quad f,g: D_i \rightarrow \mathbb{R},
\end{align*}
one obtains
\begin{align*}
\delta^\gamma \oint_{C}\psi_o \nabla \psi_o \cdot \hat{n} \; ds&= \oint_{C}(\delta^\gamma \psi_o \nabla \psi_o+\psi_o \delta^\gamma \nabla \psi_o) \cdot \hat{n} \; ds=2 \oint_{C}\psi_o  \nabla(\delta^\gamma  \psi_o) \cdot \hat{n} \; ds=-2 \oint_{C}\psi_o  (\delta^\gamma v_o \cdot \hat{t}) \; ds
\end{align*}
where $\delta^\gamma  \psi_o: D_0 \rightarrow \mathbb{R}$ is the streamfunction corresponding to the velocity field $\delta^\gamma v_o$. 
Similarly, for the integral involving $\psi_i$. Since $\delta \gamma=(\delta^\gamma v_o - \delta^\gamma v_i) \cdot t$, one finally gets
\begin{align*}
\delta^\gamma H(\vec{\Sigma},\gamma)&=-\oint_{C} \psi \delta \gamma \; ds, \\
\Rightarrow \frac{\delta h}{\delta \gamma}&=-\psi
\end{align*}
To compute $\delta h /\delta \Sigma$, use (\ref{eq:sigmafd}). Recall the functional derivatives $\delta h^o /\delta \Sigma, \delta h^i /\delta \Sigma$ in Zakharov's model are given by \cite{Za1968, Sh2016}:
\begin{align*}
\frac{\delta h^o }{\delta \Sigma}&=\frac{v_o^2}{2}-(v_o \cdot \hat{n})^2,\\
\frac{\delta h^i }{\delta \Sigma}&=-\frac{v_i^2}{2}+(v_i \cdot \hat{n})^2 \\
\Rightarrow \frac{\delta h}{\delta \Sigma}&= \frac{v_o^2-v_i^2}{2}
\end{align*}
invoking (\ref{eq:normalvel}) again.

\paragraph{Hamiltonian vector field.} The Hamiltonian vector field relative to the vortex sheet bracket is obtained from:
\begin{align*}
\dot{F}&= \{F,H \}_V, \\
\Rightarrow \oint_{C} \frac{\delta f}{\delta \Sigma} \frac{\partial \Sigma}{\partial t} \; ds+\oint_{C} \frac{\delta f}{\delta \gamma} \frac{\partial \gamma}{\partial t} \; ds&=\oint_{C} \left( -\frac{\delta f}{\delta \Sigma}\frac{\partial}{\partial s} \left( \frac{\delta h}{\delta \gamma}\right) + \frac{\delta h}{\delta \Sigma}\frac{\partial}{\partial s} \left( \frac{\delta f}{\delta \gamma}\right) \right)\;ds 
\end{align*}
which gives 
\begin{gather} \label{eq:hameqns1}
\begin{aligned}
 \frac{\partial \Sigma}{\partial t}&=-\frac{\partial}{\partial s} \left( \frac{\delta h}{\delta \gamma}\right), \\
\frac{\partial \gamma}{\partial t}&=-\frac{\partial}{\partial s} \left( \frac{\delta h}{\delta \Sigma}\right)  
\end{aligned}
\end{gather}
The final equations of motion are obtained as
\begin{gather} \label{eq:evol1}
\begin{aligned}
 \frac{\partial \Sigma}{\partial t}&=v \cdot \hat{n}=v_o \cdot    \hat{n}=v_i \cdot \hat{n}\\
\frac{\partial \gamma}{\partial t}&=-\frac{\partial}{\partial s}\left( \frac{v_{o}^2-v_{i}^2}{2}\right) 
\end{aligned}
\end{gather}
\begin{proposition} The evolution of a closed vortex sheet in a homogeneous fluid is governed by the PDE system (\ref{eq:evol1}) combined with (\ref{eq:cpv}), (\ref{eq:cpvo}) and (\ref{eq:cpvi}).  Together they form a  system of intrinsic equations for the evolution of the closed vortex sheet.  This is a Hamiltonian system of the form (\ref{eq:hameqns1}) with respect to the Poisson brackets (\ref{eq:pbvs1}) and Hamiltonian function (\ref{eq:hamB}).
  \end{proposition}
\paragraph{Remarks.} Equations (\ref{eq:evol1}) may be compared with the traditional Birkhoff-Rott evolution equation for a vortex sheet which, in the notation of this paper, is:
\begin{align*}
\frac{\partial \vec{\Sigma}}{\partial t}&=CPV_\Gamma,
\end{align*}
an integro-differential equation, where $CPV_\Gamma$ is the Cauchy Principal Value integral (\ref{eq:cpv}) written in terms of the parameter $\Gamma$ defined as $d\Gamma=\gamma d\tilde{s}$. This differential is a Lagrangian invariant of a vortex sheet evolving in a uniform density fluid. However, as Saffman  \cite{Sa1992} himself observes the choice of this Lagrangian parameter for the curve may not be the best one from a computational point of view. Moreover, the Lagrangian invariance of this differential is lost when the sheet separates fluids of different densities, the case considered in the next section.  This is discussed more in the next section.

\subsection{The case of a density field with a jump.} For this case, assume there is non-trivial flow in both domains with unequal fluid densities $\rho_o$ and $\rho_i$ in the outer and inner domains, respectively. The dynamical variables are now all correspondingly density weighted. 
\begin{align}
\tilde{\gamma}&=(\tilde{v}_{o}-\tilde{v}_{i})\cdot \hat{t}, \nonumber \\
&=:(\rho_o{v}_{o}-\rho_i {v}_{i})\cdot \hat{t} \quad \rho_o \neq \rho_i \label{eq:assumptions2}
\end{align}
It may be noted that $\tilde{v}_{o}$ and $\tilde{v}_{i}$ are {\it also vorticity-free}.
The following relations continue to hold:
\begin{align}
v_{o}&=\frac{\gamma}{2} \hat{t}+CPV \nonumber \\
v_{i}&=-\frac{\gamma}{2}\hat{t}+CPV \nonumber \\
v_{o}\cdot \hat{n}&=v_{i}\cdot \hat{n}=CPV \cdot \hat{n} \label{eq:normalvel2}
\end{align}
The velocity potential variables are now:
\begin{align*}
\tilde{\phi}_{o}&:=\rho_o \phi_{o}=\tilde{\Phi}_{o} \mid_{C}, \quad  \tilde{\Phi}_{o}:=\rho_o\Phi_{o}, \\
\tilde{\phi}_{i}&:=\rho_i \phi_{i}=\tilde{\Phi}_{i} \mid_{C}, \quad  \tilde{\Phi}_{i}:=\rho_i\Phi_{i}
\end{align*}
Relation (\ref{eq:normalvelphi}) holds unchanged, but (\ref{eq:phigammasi}) now holds for the tilde variables, i.e.
\begin{align}
\delta \tilde{\gamma}&=\delta \left(\frac{\partial (\tilde{\phi}_{o}-\tilde{\phi}_{i})}{\partial s} \right)=\frac{\partial (\delta (\tilde{\phi}_{o}-\tilde{\phi}_{i}))}{\partial s} \label{eq:phigammatilde}
\end{align}
The total variation of $F(\vec{\Sigma},\tilde{\gamma})$ must be the same as the total variation of $F_o(\vec{\Sigma},\tilde{\phi}_{o})$ and $F_i(\vec{\Sigma},\tilde{\phi}_{i})$, i.e.
\begin{align}
&\oint_{C} \frac{\delta f}{\delta \Sigma} \delta \Sigma \; ds +\oint_{C} \frac{\delta f}{\delta \tilde{\gamma}} \delta \tilde{\gamma}\; ds  \nonumber \\
&\hspace{1in} =\oint_{C} \frac{\delta f_o}{\delta \Sigma} \delta \Sigma \; ds+ \oint_{C} \frac{\delta  f_i}{\delta \Sigma} \delta \Sigma \; ds+\oint_{C} \frac{\delta f_o}{\delta \tilde{\phi}_{o}} \delta \tilde{\phi}_{o} \; ds  +\oint_{C} \frac{\delta f_i}{\delta \tilde{\phi}_{i}} \delta \tilde{\phi}_{i} \; ds \label{eq:totalsidens}
\end{align}
Substituting (\ref{eq:phigammatilde}), the total variation equation becomes
\begin{align*}
&\oint_{C} \frac{\delta f}{\delta \Sigma}\delta \Sigma \; ds -\oint_{C} \frac{\partial}{\partial s} \left(\frac{\delta f}{\delta \tilde{\gamma}}\right) \delta \tilde{\phi}_{o} \; ds + \oint_{C} \frac{\partial}{\partial s} \left(\frac{\delta f}{\delta \tilde{\gamma}}\right) \delta \tilde{\phi}_{i} \; ds  \\
&\hspace{1in} =\oint_{C} \frac{\delta f_o}{\delta \Sigma} \delta \Sigma \; ds+ \oint_{C} \frac{\delta f_i}{\delta \Sigma} \delta \Sigma \; ds+\oint_{C} \frac{\delta f_o}{\delta \tilde{\phi}_{o}} \delta \tilde{\phi}_{o} \; ds  +\oint_{C} \frac{\delta f_i}{\delta \tilde{\phi}_{i}} \delta \tilde{\phi}_{i} \; ds  
\end{align*}

This leads to the following relations between the functional derivatives:
\begin{align}
\frac{\delta f_o}{\delta \Sigma}+\frac{\delta f_i}{\delta \Sigma}&=\frac{\delta f}{\delta \Sigma}   \label{eq:sigmafdtilde} \\
\frac{\delta f_o}{\delta \tilde{\phi}_{o}} &=-\frac{\partial}{\partial s} \left( \frac{\delta f}{\delta \tilde{\gamma}}\right)
 \label{eq:phigammafdsi1tilde} \\
\frac{\delta f_i}{\delta \tilde{\phi}_{i}} &=\frac{\partial}{\partial s} \left( \frac{\delta f}{\delta \tilde{\gamma}}\right), \label{eq:phigammafdsi2tilde}
\end{align}
From (\ref{eq:phigammafdsi1tilde}) and (\ref{eq:phigammafdsi2tilde}), we get 
\begin{align}
\frac{\delta f_o}{\delta \tilde{\phi}_{o}} +
\frac{\delta f_i}{\delta \tilde{\phi}_{i}}&=0 \label{eq:reltilde}
\end{align}

\paragraph{Vortex sheet bracket.} The Zakharov bracket we consider is now the sum of two Zakharov brackets in the density-weighted variables $(\Sigma, \tilde{\phi}_{so},\tilde{\phi}_{si})$: 
\begin{align*}
\left\{F,G \right\}_Z&:=\oint_{C}\left( \frac{\delta f_o}{\delta \Sigma } \frac{\delta g_o}{\delta \tilde{\phi}_{o}} -\frac{\delta f_o}{\delta \tilde{\phi}_{o}} \frac{\delta g_o}{\delta \Sigma }\right)  \, ds - \oint_{C}  \left(\frac{\delta f_i}{\delta \Sigma } \frac{\delta g_i}{\delta \tilde{\phi}_{i}} - \frac{\delta f_i}{\delta \tilde{\phi}_{i}} \frac{\delta g_i}{\delta \Sigma } \right)  \, ds   \\
&=\oint_{C}\left(\left( \frac{\delta f_o}{\delta \Sigma } +\frac{\delta f_i}{\delta \Sigma } \right)\frac{\delta g_o}{\delta \tilde{\phi}_{o}} -\frac{\delta f_o}{\delta \tilde{\phi}_{o}}\left( \frac{\delta g_o}{\delta \Sigma }+\frac{\delta g_i}{\delta \Sigma } \right)\right)  \, ds, 
\end{align*}
using (\ref{eq:reltilde}). 

Substituting the relations between the functional derivatives one gets
\begin{align*}
\left\{F,G \right\}_Z&=\oint_{C} \left[\left( \frac{\delta f}{\delta \Sigma}
\right) \left(-\frac{\partial}{\partial s} \left( \frac{\delta g}{\delta \gamma}\right) \right) - \left( \frac{\delta g}{\delta \Sigma}\right) \left(-\frac{\partial}{\partial s} \left( \frac{\delta f}{\delta \gamma}\right) \right)\right]\, ds 
\end{align*}
which leads to the same Poisson bracket as for the uniform density case, i.e.
\begin{align}
\left\{F,G\right\}_V(\vec{\Sigma},\tilde{\gamma})&=\oint_{C} \left[-\frac{\delta f}{\delta \Sigma} \frac{\partial}{\partial s} \left( \frac{\delta g}{\delta \tilde{\gamma}}\right)+  \frac{\delta g}{\delta \Sigma} \frac{\partial}{\partial s} \left( \frac{\delta f}{\delta \tilde{\gamma}}\right)\right]\, ds \label{eq:pbvs2}
\end{align}
and 
\begin{gather} \label{eq:hameqns2}
\begin{aligned}
 \frac{\partial \Sigma}{\partial t}&=-\frac{\partial}{\partial s} \left( \frac{\delta h}{\delta \tilde{\gamma}}\right), \\
\frac{\partial \tilde{\gamma}}{\partial t}&=-\frac{\partial}{\partial s} \left( \frac{\delta h}{\delta \Sigma}\right)  
\end{aligned}
\end{gather}
for the Hamiltonian function which is now the sum of the kinetic energy and {\it the potential energy.} 
\paragraph{Hamiltonian function and functional derivatives.} The Hamiltonian function is the total energy to which the potential energy also makes a contribution. 
To obtain the potential energy expression, we follow the derivation in \cite{Sh2016}:
\begin{align*}
P.E.&:=- \rho_o \int_{D_o} \vec{g}_r \cdot y \hat{j}  \; \mu - \rho_i \int_{D_i} \vec{g}_r \cdot y \hat{j} \; \mu
\end{align*}
where $\vec{g}_r$ is the constant gravity vector in the direction $-\hat{j}$ and $y$ is the height relative to a datum.  Using standard vector identities, rewrite the potential energy terms as
\begin{align*}
P.E.&=- \rho_o  \vec{g}_r \cdot \int_{D_o}  \nabla \frac{y^2}{2}  \; \mu  -   \rho_i \vec{g}_r  \cdot \int_{D_i} \nabla \frac{y^2}{2} 
 \; \mu, \\
&=   -\rho_o  \vec{g}_r \cdot \oint_C   \frac{y^2}{2}\; \hat{n}  \;  ds  -\rho_o  \vec{g}_r \cdot \oint_{C_\infty}   \frac{y^2}{2}\; \hat{n}  \;  ds +  \rho_i \vec{g}_r \cdot  \oint_{C}    \frac{y^2}{2}\; \hat{n}  \;  ds, 
\end{align*}
recalling that $\hat{n}$ is outward pointing on $D_o$ and inward pointing on $D_i$. In the above, $C_\infty \in D_o$, is any outer boundary that completely encloses $D_i$ and goes to infinity. The contour integral can be easily shown to be zero by taking, for example, a rectangular boundary whose lengths go to infinity, and noting that the individual integrals on the top and bottom sides/left and right sides cancel.   And so 
\begin{align*}
P.E.&=(\rho_i -\rho_o ) \vec{g}_r \cdot \oint_C   \frac{y^2}{2}\; \hat{n } \;  ds 
\end{align*}
Variations in $\Sigma$ will cause variations in the above integral and make contributions to the functional derivative of the P.E..  

The variations $\delta y$ and $\delta \Sigma$ are related by 
\begin{align*}
\delta y \hat{j} \cdot n &=\delta \Sigma,
\end{align*}
and so the variation in each of the two integrals takes the form:
\begin{align*}
\rho_o \vec{g}_r \cdot \oint_C  y \delta y  \; \hat{n} \; ds= - \rho_o g_r  \oint_C y \delta \Sigma  \; ds \\
\rho_i \vec{g}_r \cdot \oint_C  y \delta y  \; \hat{n} \; ds= - \rho_i g_r  \oint_C y \delta \Sigma  \; ds 
\end{align*}
The P.E. term is the same in the Zakharov model and one thinks of it as the sum of two P. E. terms, one for  each domain. This leads to the following functional derivatives:
\begin{align*}
\frac{\delta (P.E.)_o}{\delta \Sigma}&=\rho_o g_r  y, \quad \frac{\delta (P.E.)_i}{\delta \Sigma}=-\rho_i g_r  y
\end{align*}
The Hamiltonian function is 
\begin{align}
H(\vec{\Sigma},\gamma)&:=\oint_C h_o \; ds + \oint_C h_i \; ds \nonumber \\
&=\oint_{C}\left(-\frac{\rho_o}{2} \psi \tilde{v}_{o} \cdot \hat{t} - \rho_o  \vec{g}_r \cdot  \frac{y^2}{2}\; \hat{n} \right) \; ds+  \oint_{C}\left(\frac{\rho_i}{2} \psi \tilde{v}_{i} \cdot \hat{t} - \rho_i \vec{g}_r \cdot  \frac{y^2}{2}\; \hat{n} \right) \; ds \nonumber \\
&= -\frac{1}{2} \oint_{C} \psi  \tilde{\gamma} \; ds  + (\rho_i -\rho_o ) \vec{g}_r \cdot \oint_C   \frac{y^2}{2}\; \hat{n}  \;  ds \label{eq:ham2}
\end{align}

 The functional derivatives of $H$ are obtained as 
\begin{align*}
\frac{\delta h}{\delta \tilde{\gamma}}&=-\psi, \\
\frac{\delta h}{\delta \Sigma}&=\frac{\delta h_o}{\delta \Sigma}+\frac{\delta h_i}{\delta \Sigma}  \\
&=\rho_o \left(\frac{ v_{o}^2}{2}+ g_r y-(v_o \cdot \hat{n})^2\right)-\rho_i \left(\frac{ v_{i}^2}{2} - g_r y -(v_i \cdot n)^2 \right),
\end{align*}
the functional derivatives in Zakharov's model being the ones in the uniform density model weighted by the respective densities. 
 
 The final equations of motion are obtained as
\begin{gather} \label{eq:evol2}
\begin{aligned}
 \frac{\partial \Sigma}{\partial t}&=v \cdot \hat{n}=v_o \cdot \hat{n}=v_i \cdot  \hat{n} \\
\frac{\partial \tilde{\gamma}}{\partial t}&=-\frac{\partial}{\partial s}\left( \frac{\rho_o v_{o}^2}{2}-\frac{\rho_i v_{i}^2}{2}+ (\rho_o-\rho_i)(g_r y -(v \cdot \hat{n})^2)
\right) 
\end{aligned}
\end{gather}
\begin{proposition} The evolution of a closed vortex sheet separating two fluids of piecewise constant density is governed by the PDE system (\ref{eq:evol1}) combined with (\ref{eq:cpv}), (\ref{eq:cpvo}) and (\ref{eq:cpvi}).  Together they form an intrinsic system of equations for the evolution of the closed vortex sheet.  This is a Hamiltonian system of the form (\ref{eq:hameqns2}) with respect to the Poisson brackets (\ref{eq:pbvs2}) and Hamiltonian function (\ref{eq:ham2}).
  \end{proposition}
System (\ref{eq:evol2}) reduces to (\ref{eq:evol1}) when $\rho_o=\rho_i$. 

\paragraph{Remarks.} The Poisson brackets (\ref{eq:pbvs1}) and (\ref{eq:pbvs2}) bear resemblance to the KdV bracket and its generalizations and to (at least a part of) the compressible fluid flow bracket \cite{MoGr1980, Mo1981, DuNo1983, MoFe1990, MaRa1999}. These connections promise to be interesting but are not explored in this paper. 

\paragraph{Obtaining the $\tilde{\gamma}$ equation in (\ref{eq:evol2}) from Euler's equation.} In attempting to derive (\ref{eq:evol2}) from Euler's equations, it is important to make the following distinction when expressing the time derivative $\partial(\;) / \partial t$ at a point. In the former, the underlying point is on a moving curve whereas in the latter the underlying point is fixed in the spatial domain.  This is analogous to the distinction between $\partial \phi/\partial t$ and $\partial \Phi /\partial t$ in Zakharov's formulation \cite{Za1968}; see also \cite{Mi1996}.

%To keep this distinction in mind, introduce sheet strength variables, with subscript $S$, viewed as functions of domain points $r(s,t) \in C \subset \mathbb{R}^2$ and $t$:
%\begin{align*}
%\tilde{\gamma}_S(r(s),t)&:=(\rho_o{v}_{o}(r(s),t)-\rho_i {v}_{i}(r(s),t))\cdot \hat{t}(r(s),t) \\
%\gamma_S(r(s),t)&:=({v}_{o}(r(s),t)-{v}_{i}(r(s),t))\cdot \hat{t}(r(s),t)
%\end{align*}

First, we note the following relation obtained by using previous definitions and relations:
\begin{align}
\tilde{\gamma}&=\rho_o \left(\frac{\gamma}{2}+ CPV \cdot \hat{t}\right)-\rho_i\left(-\frac{\gamma}{2}+ CPV \cdot \hat{t}\right) \nonumber \\
&=\left(\frac{\rho_o+\rho_i}{2} \right)\gamma+ \left(\rho_o-\rho_i \right)CPV \cdot\hat{t} \label{eq:ggtilde}
\end{align}
Next, from the Euler's equations written for the outer and inner velocity fields, obtain the following equation. Applied on $C$, assuming pressure continuity across $C$, this gives: 
\begin{align*}
\rho_o \frac{Dv_{o}}{Dt} \cdot \hat{t}-\rho_i \frac{Dv_{i}}{Dt} \cdot \hat{t}&=(\rho_i -\rho_0) g_r \hat{j}\cdot \hat{t}
\end{align*}
Substituting for the total derivatives, using the fact that domains are vorticity-free, and using the relations above (\ref{eq:normalvel2}), gives:
\begin{align}
&\rho_o \left(\frac{\partial v_{o}}{\partial t}_{\mid_S }\cdot \hat{t}+ \frac{\partial}{\partial s} \left( \frac{v_{o}^2}{2}\right)\right)-\rho_i \left(\frac{\partial v_{i}}{\partial t}_{\mid_S } \cdot \hat{t}+ \frac{\partial}{\partial s} \left( \frac{v_{i}^2}{2}\right)\right)=(\rho_i -\rho_0) g_r \hat{j}\cdot \hat{t} \nonumber \\
&\Rightarrow \rho_o \left(\frac{\partial}{\partial t}_{\mid_S }\left( \frac{\gamma \hat{t}}{2} +CPV \right)\cdot \hat{t}+ \frac{\partial}{\partial s} \left( \frac{v_{o}^2}{2}\right)\right) \nonumber \\
&\hspace{1in}-\rho_i \left(\frac{\partial}{\partial t}_{\mid_S }\left( -\frac{\gamma \hat{t}}{2} +CPV \right)\cdot \hat{t}+ \frac{\partial}{\partial s} \left( \frac{v_{i}^2}{2}\right)\right)=(\rho_i -\rho_0) g_r \hat{j}\cdot \hat{t} \nonumber \\
&\Rightarrow \left(\frac{\rho_o +\rho_i}{2}\right)\left(\frac{\partial \gamma}{\partial t}_{\mid_S } \hat{t} \cdot \hat{t} + \frac{\gamma}{2}\frac{\partial \hat{t}}{\partial t}_{\mid_S } \cdot  \hat{t} \right)+(\rho_o-\rho_i)\frac{\partial (CPV)}{\partial t}_{\mid_S } \cdot \hat{t} \nonumber \\
&\hspace{2in}+ \frac{\partial}{\partial s} \left( \frac{\rho_o v_{o}^2}{2}-\frac{\rho_i v_{i}^2}{2}\right)=(\rho_i -\rho_o) g_r \hat{j}\cdot \hat{t} \nonumber \\
&\Rightarrow \left(\frac{\rho_o +\rho_i}{2}\right)\frac{\partial \gamma}{\partial t}_{\mid_S } +(\rho_o-\rho_i)\frac{\partial (CPV)}{\partial t}_{\mid_S } \cdot \hat{t}+ \frac{\partial}{\partial s} \left( \frac{\rho_o v_{o}^2}{2}-\frac{\rho_i v_{i}^2}{2}\right)=(\rho_i -\rho_o) g_r \hat{j}\cdot \hat{t} \nonumber \\
&\Rightarrow \frac{\partial \tilde{\gamma}}{\partial t}_{\mid_S }-(\rho_o-\rho_i) (CPV)\cdot\frac{\partial \hat{t}}{\partial t} _{\mid_S }+ \frac{\partial}{\partial s} \left( \frac{\rho_o v_{o}^2}{2}-\frac{\rho_i v_{i}^2}{2}\right)=(\rho_i -\rho_o) g_r \hat{j}\cdot \hat{t}   \label{eq:gammatilevol}
\end{align}
where we have invoked (\ref{eq:ggtilde}) and used  the elementary fact that  $\partial{(\hat{t} \cdot \hat{t})}/\partial t_{\mid_S}=0=2 \hat{t} \cdot (\partial \hat{t}/\partial t)_{\mid_S}$. The notation $_{\mid_S}$ is used to denote time derivative evaluated at a spatial point. 

\begin{figure}
\centering
\vspace{-2in}
\includegraphics[width=6in]{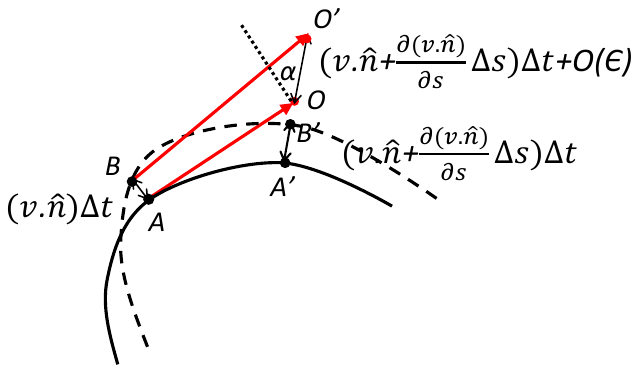}
\vspace{-4in}
\caption{Sketch illustrating the computation of the $\partial \hat{t}/\partial t_{\mid_S}$ term. }
\label{tan}
\end{figure}

   To obtain an expression for $\partial \hat{t}/\partial t_{\mid_S }$ due to the normal displacement of $C$, consider the schematic sketch in Figure \ref{tan}. 
The curve $C$ is shown in solid black at time $t$  and in dashed black at time $t + \triangle t$. Points $A$ and $B$ denote two close points on $C$ at time $t$, separated by $\triangle s$. Consider the infinitesimal line element of length $\triangle l$, shown in red, connecting points $A$ and $O$, and parallel to $\hat{t}$.  The point $O \notin C$, and is located at a distance $\epsilon$ from $B$ along the normal direction. At time $t+\triangle t$, the updated positions are $A',B'$ and $O'$, due to normal displacements by the amounts shown in the figure. It follows that $O$ has $AO^\perp$ and $AO^\parallel$ velocity components, relative to $A$, given by 
\[\left(\frac{\partial (v \cdot \hat{n})}{\partial s}_{\mid_A} \triangle s +O(\epsilon) \right)\cos \alpha, \quad \left(\frac{\partial (v \cdot \hat{n})}{\partial s}_{\mid_A} \triangle s +O(\epsilon)\right)\sin \alpha,\]
respectively. Now, as $\triangle t \rightarrow 0, \triangle s \rightarrow 0$, \[\triangle l \rightarrow \triangle s, \alpha \rightarrow 0, \epsilon\rightarrow 0, \hat{n}(s+\triangle s) \rightarrow \hat{n}(s),\] leading to a rotation rate of segment $AO$ whose value is given by 
\[\frac{\partial (v \cdot \hat{n})}{\partial s}_{\mid_A},\]
and to the final expression
\begin{align*}
\frac{\partial \hat{t}}{\partial t}_{\mid_S}&=\frac{\partial (v \cdot \hat{n})}{\partial s}\hat{n}.
\end{align*}
\paragraph{Remark.} No curvature term appears in the above equation. This is due to the framework of this model, mentioned earlier, in which $C$ is advanced only by normal displacements. If tangential displacements of $C$ were considered, as is the case when using a Lagrangian parameter, the tangential displacements of both points $B$ and $B'$ in Figure \ref{tan} would lead to a curvature term.  A curvature terms appears, for example, in viscous formulations of free surface dynamics with either the zero-shear-stress condition or the pressure conditions applied at the interface, since these formulations advance the interface as a material surface \cite{Ba1967, LuKo1999, BrThLeHo2014}. \\

Substituting this in (\ref{eq:gammatilevol}) and using (\ref{eq:normalvel2}), one obtains
\begin{align*}
&\frac{\partial \tilde{\gamma}}{\partial t}_{\mid_S}+ \frac{\partial}{\partial s} \left( \frac{\rho_o v_{o}^2}{2}-\frac{\rho_i v_{i}^2}{2}-(\rho_o-\rho_i)\frac{(v \cdot n)^2}{2} \right)=(\rho_i -\rho_o) g_r \hat{j}\cdot t,
\end{align*}
The time derivative with respect to a spatial point and a point on the moving curve differ, respectively, by the
convective acceleration due to the normal sheet velocity, 
\begin{align*}
\frac{\partial \tilde{\gamma}}{\partial t}&=\frac{\partial \tilde{\gamma}}{\partial t}_{\mid_S}+\frac{\partial \Sigma}{\partial t} \hat{n} \cdot \nabla \tilde{\gamma} \\
&=\frac{\partial \tilde{\gamma}}{\partial t}_{\mid_S}+(v\cdot n) \hat{n} \cdot  \left( \nabla(\tilde{v}_o-\tilde{v}_i) \cdot \hat{t}+ (\tilde{v}_o-\tilde{v}_i)  \cdot \nabla \hat{t} \right) \\
&=\frac{\partial \tilde{\gamma}}{\partial t}_{\mid_S}+(v\cdot n) \hat{n} \cdot  \left(\nabla(\tilde{v}_o-\tilde{v}_i) \cdot \hat{t} \right),
\end{align*}since $
(\tilde{v}_o-\tilde{v}_i)  \cdot \nabla \hat{t} =0,$
 again due to the fact that curve-tangent displacements are not considered in this framework. 

   It can be shown easily that for a velocity field in domain $D$,
\begin{align*}
\hat{n} \cdot  \left(\nabla v \cdot \hat{t} \right)_{\mid_C}&=\nabla \left( v \cdot \hat{n} \right) \cdot  \hat{t}_{\mid_C}
\end{align*}
Start with a local rectangular coordinate system $(n_r,s_r)$ based at $p \in C$ and spanned by constant unit vectors $(\hat{n}_r,\hat{t}_r)$ that are equal to $(\hat{n},\hat{t})$ at $p$. It follows that 
\begin{align*}
\frac{\partial f}{\partial s_r}_{\mid_p}&=\frac{\partial f}{\partial s}_{\mid_p}, \quad f: C \rightarrow \mathbb{R}
\end{align*}
In these coordinates, 
\begin{align*}
\nabla v(p)&= \left(\begin{array}{cc} \frac{\partial (v\cdot \hat{n}_r)}{\partial n_r} & \frac{\partial (v \cdot \hat{n}_r)}{\partial s_r} \\
\frac{\partial (v\cdot \hat{t}_r)}{\partial n_r} & \frac{\partial (v \cdot \hat{t}_r)}{\partial s_r}
\end{array}\right)_{\mid_p} \\ 
\Rightarrow \nabla v \cdot \hat{t}_r&=\frac{\partial (v \cdot \hat{n}_r)}{\partial s_r}\hat{n}_r+ \frac{\partial (v \cdot \hat{t}_r)}{\partial s_r}\hat{t}_r \\
\Rightarrow \hat{n}_r \cdot \left(\nabla v \cdot \hat{t}_r\right)&=\frac{\partial (v \cdot \hat{n}_r)}{\partial s_r} \\
\Rightarrow \hat{n} \cdot \left(\nabla v \cdot \hat{t} \right)&=\frac{\partial (v \cdot \hat{n})}{\partial s} 
\end{align*}
The result is trivially extended to the weighted velocities. 
Finally, using (\ref{eq:normalvel2}), 
\begin{align*}
\frac{\partial \tilde{\gamma}}{\partial t}&=\frac{\partial \tilde{\gamma}}{\partial t}_{\mid_S}+(v\cdot n)  \frac{\partial   \left((\tilde{v}_o-\tilde{v}_i) \cdot \hat{n} \right)}{\partial s} \\
&=\frac{\partial \tilde{\gamma}}{\partial t}_{\mid_S}+(\rho_o-\rho_i)\frac{\partial}{\partial s} \left(\frac{(v \cdot n)^2}{2} \right),
\end{align*}
which leads to the same evolution equation (\ref{eq:evol2}) for the intrinsic sheet strength $\tilde{\gamma}$.  

%The reader is referred to the cited literature for alternative derivations of the evolution equations.

\section{Lagrangian invariants of the sheet motion.}

  In this section,  Lagrangian invariants of the sheet motion will be discussed, where the convecting velocity field is the $CPV$ field of the vortex sheet.

    Recall that the Cauchy Principal Value (\ref{eq:cpv}) is the velocity of any material point $p \in C$, and will be denoted by $\bar{v}$ in this section,
\begin{align}
\bar{v}&=\frac{v_{o}+v_{i}}{2}_{\mid_C }\label{eq:barv}
\end{align}
The $CPV$ field is therefore a velocity field defined only on the sheet. The `bar' notation will also be used to denote operators and quantities associated with the $CPV$ field. Consider now the Lagrangian transport of the following differential
\[v \cdot d \mathbf{s}, \quad d\mathbf{s}=\hat{t} ds\] by the $CPV$ field, where $v$ could represent either $v_o$ or $v_i$: : 
\begin{align*}
\frac{\bar{D}(v \cdot d \mathbf{s})}{Dt}&=\frac{\bar{D}v}{Dt}\cdot d\mathbf{s}+v \cdot \frac{\bar{D}(d \mathbf{s})}{Dt} \\ 
&=\left(\frac{\partial v}{\partial t}+ \bar{v}\cdot \nabla v\right)\cdot d \mathbf{s}+v \cdot (\nabla \bar{v} \cdot d \mathbf{s})
\end{align*}
where we have used the standard formula for the rate of change of infinitesimal material line elements \cite{Ba1967}:
\begin{align*}
\frac{D (d \mathbf{s})}{Dt}&=\nabla v \cdot d{\mathbf{s}},
\end{align*}
Let
\begin{align*}
\nabla v=:M, \quad \nabla \bar{v}=:\bar{M}
\end{align*} 
denote deformation matrices. In the definition of $\bar{M}$, $\nabla$ is viewed as a one-sided operator on $C$--the side determined by the domain of the velocity field it acts on,  
\begin{align*}
\bar{M}&=\frac{\nabla_o v_o + \nabla_i v_i}{2}_{\mid_C }=:\frac{1}{2} \left(M_{o}+M_{i} \right),
\end{align*}
where $M_{o}:=\nabla_o v_{o}$ and $M_{i}=\nabla_o v_{i}$. Due to the flow being irrotational on either side of the sheet, both deformation matrices are symmetric, i.e. 
\begin{align*}
M_{i}^T&=M_{i}, \quad M_{o}^T=M_{o},
\end{align*}
In terms of these matrices,  
\[(\bar{v}\cdot \nabla v) \cdot d \mathbf{s} \equiv \bar{v}^T\cdot  M^T \cdot d \mathbf{s}, \quad  v\cdot (\nabla \bar{v} \cdot d \mathbf{s}) \equiv v^T\cdot  \bar{M} \cdot d \mathbf{s}\]
(with $\cdot$ denoting matrix multiplication and $d\mathbf{s}$ a column vector, on the right), and so
\begin{align*}
\frac{\bar{D}(v \cdot d \mathbf{s})}{Dt}&=\left(\frac{\partial v^T}{\partial t}+ \bar{v}^T\cdot M^T+v^T \cdot \bar{M}  \right)\cdot d \mathbf{s}
\end{align*}

\subsection{Vortex sheet in a homogeneous fluid.}
It follows that  on $C$,
\begin{align*}
&\frac{\bar{D}(\gamma ds)}{Dt}\\
&=\frac{\bar{D}(v_o \cdot d \mathbf{s})}{Dt}-\frac{\bar{D}(v_i \cdot d \mathbf{s})}{Dt}\\
&=\left(\frac{\partial v_{o}^T}{\partial t}+ \bar{v}^T\cdot M_{o}^T+v_{o}^T \cdot \bar{M} - \frac{\partial v_{i}^T}{\partial t}- \bar{v}^T\cdot M_{i}^T-v_{i}^T \cdot \bar{M}\right)\cdot d \mathbf{s} \\
&=\left(\frac{\partial v_{o}^T}{\partial t}+ \frac{v_{o}^T+v_{i}^T}{2}\cdot M_{o}^T+v_{o}^T \cdot \frac{1}{2} \left(M_{o}+M_{i} \right) \right. \\
& \left. \hspace{1in}  - \frac{\partial v_{i}^T}{\partial t}- \frac{v_{o}^T+v_{i}^T}{2}\cdot M_{i}^T-v_{i}^T \cdot\frac{1}{2} \left(M_{o}+M_{i} \right) \right)\cdot d \mathbf{s}  \\
&=\left(\frac{D v_{o}^T}{D t}- \frac{D v_{i}^T}{D t} \right)\cdot d \mathbf{s}  \\
&=\left(\frac{\nabla p_{o}}{\rho}- \frac{\nabla p_{i}}{\rho} \right)\cdot d \mathbf{s},  \\
&=0,
\end{align*}
applying pressure continuity across the sheet, leading to to the well-known fact that $\gamma ds$ is a Lagrangian invariant \cite{Sa1992} for a homogeneous vortex sheet. But for unequal material densities on either side, it no longer is. 

 However, 
\begin{align*}
&\frac{\bar{D}(v_o \cdot d \mathbf{s})}{Dt}\\
&=\left(\frac{\partial v_{o}^T}{\partial t}+ \bar{v}^T\cdot M_{o}^T+v_{o}^T \cdot \bar{M} \right)\cdot d \mathbf{s} \\
&=\left(\frac{\partial v_{o}^T}{\partial t}+ \frac{v_{o}^T+v_{i}^T}{2}\cdot M_{o}^T+v_{o}^T \cdot \frac{1}{2} \left(M_{o}+M_{i}  \right) \right)\cdot d \mathbf{s}  \\
&=\left(\frac{D v_{o}^T}{D t}+\frac{1}{2}\left( v_{o} \cdot \nabla_i v_{i}+v_{i} \cdot \nabla_o v_{o}\right)\right)\cdot d \mathbf{s}   \\
&=\left(-\frac{\nabla p_{o}}{\rho}+\vec{g}+\frac{1}{2}\left( v_{o} \cdot \nabla_i v_{i}+v_{i} \cdot \nabla_o v_{o}\right)\right)\cdot d \mathbf{s}
\end{align*}
To evaluate the second and third terms, invoke again the local rectangular coordinate system (after (\ref{eq:gammatilevol})), making use again of one-sided derivatives and the fact that $\nabla_i v_{i}$ and $\nabla_o v_{o}$ are  symmetric, to obtain
\begin{align*}
(v_{o} \cdot \nabla_i v_{i})\cdot \hat{t}&=v_{no}\frac{ \partial v_{ti}}{\partial n}+ v_{to}\frac{ \partial v_{ti}}{\partial s}=v_{no}\frac{ \partial v_{ni}}{\partial s}+ v_{to}\frac{ \partial v_{ti}}{\partial s} \\
(v_{i} \cdot \nabla_o v_{o})\cdot \hat{t}&=v_{ni}\frac{ \partial v_{to}}{\partial n}+ v_{ti}\frac{ \partial v_{to}}{\partial s}=v_{ni}\frac{ \partial v_{no}}{\partial s}+ v_{ti}\frac{ \partial v_{to}}{\partial s}
\end{align*}
Finally, we get 
\begin{align*}
\frac{\bar{D}(v_o \cdot d \mathbf{s})}{Dt}&=
\frac{\partial }{\partial s} \left(-\frac{p_{o}}{\rho}-g_r y+\frac{v_o \cdot v_i}{2}\right)ds, 
\end{align*}
leading to 
\begin{align}
\frac{\bar{d}}{dt}\oint_{C(s,t)} v_o \cdot d \mathbf{s}&=0 \label{eq:volag}
\end{align}
where $\bar{d}/{dt}$ reperesents rate of change of the controur integral convected by the $CPV$ field. The following is therefore also true: 
\begin{align}
\frac{\bar{d}}{dt} \oint_{C(s,t)} v_i \cdot d \mathbf{s}&=0 \label{eq:vilag}
\end{align}

%For the term $\nabla\left( v_{o} \cdot v_{i} \right)$ to be well-defined, make use of local extensions of these fields across $C$, $v_o^i$ and $v_i^o$, satisfying the following on $C$:
%\[v_o^i=v_o, \; \nabla v_o^i=\nabla v_o, \quad v_i^o=v_i, \; \nabla v_i^o=\nabla v_i\]
%
%
%
%However, in such a case, there exists global invariants on the sheet, as shown below.

\subsection{Vortex sheet in a fluid with a density jump.}
It is obvious that (\ref{eq:volag}) and (\ref{eq:vilag}) continue to hold for the inhomogeneous problem, with $\rho$ replaced by $\rho_o$ and $\rho_i$, respectively, and therefore 
\begin{align*}
\frac{\bar{d}}{dt}\oint_{C(s,t)} \tilde{\gamma}ds&=\frac{\bar{d}}{dt}\oint_{C(s,t)} \left(\rho_o  v_o- \rho_i v_i\right) \cdot d \mathbf{s}=\oint_{C(s,t)}\frac{\partial }{\partial s} \left((\rho_o-\rho_i)(-g_r y+\frac{v_o \cdot v_i}{2})\right)ds=0
\end{align*}
In conclusion, for the inhomogeneous problem neither of the differentials $\gamma ds$ or $\tilde{\gamma} ds$ is a Lagrangian invariant of the $CPV$ field, but the associated contour integrals  $\oint_{C(s,t)}\gamma ds$ and $\oint_{C(s,t)}\tilde{\gamma} ds$ are Lagrangian invariants of the $CPV$ field.

\paragraph{Relation to Kelvin's Circulation Theorem (KCT).} These invariants are strongly suggestive of KCT, but there is a technical difference, namely, that the convecting velocity field is different from the velocity field in the `$v \cdot d\mathbf{s}$' term. Denoting by $d^o/dt$ and $d^i/dt$, the time rate of change for convection by $v_o$ and $v_i$, respectively, KCT gives, following a similar calculation procedure, 
\[\frac{d^o}{dt}\oint_{C(s,t)} v_o \cdot d\mathbf{s}=0, \quad \frac{d^i}{dt}\oint_{C(s,t)} v_i \cdot d\mathbf{s}=0, \quad \frac{\bar{d}}{dt}\oint_{C(s,t)} CPV \cdot d\mathbf{s}=0.\]
The last equation implies that the sum of (\ref{eq:volag}) and (\ref{eq:vilag}) is zero, but does not lead to either of these equations.

\begin{proposition} In two ideal fluids of densities $\rho_o$ and $\rho_i$ separated by a closed vortex sheet, the following is an integral invariant of the sheet motion convected by the $CPV$ velocity field.
\begin{align*}
\oint_{C(s,t)}\tilde{\gamma}\; ds, 
\end{align*}
\end{proposition}
\paragraph{Remarks.} The lack of local invariance of $\gamma ds$ for the inhomogeneous problem may be related to the baroclinic generation of vorticity at the sheet (without diffusion into the domain).

\section{Future directions.} 

%This paper revisits vortex sheet evolution equations, focusing on the configuration of a closed vortex sheet in the plane separating two incompressible fluids with different constant densities, in the absence of viscous effects and surface tension. The model has potential applications to bubble dynamics. Viewing the problem as the sum of two Zakharov problems, Poisson brackets, containing the curve-tangential operator $\partial / \partial s$, are obtained for the intrinsic evolution equations. This bracket structure for sheet evolution appears to be new.  A Lagrangian invariant for this model is also derived.

     The Hamiltonian formalism presented in this paper could in principle be extended to the model considered in \cite{ShKi2021}. In that model, the domain $D_i$ has uniform vorticity everywhere and the configuration was viewed as a buoyant vortex patch of density $\rho_i$ surrounded by irrotational fluid of density $\rho_o$, requiring a vortex sheet at the patch boundary to enforce pressure continuity. The vorticity in the patch is passively convected (`contour dynamics' \cite{ZaHuRo1979}), but the Hamiltonian function will include a domain integral representing the contribution of the vortex patch to the total flow kinetic energy.

  The PDEs in this paper require further investigation using functional analysis tools and numerical simulations. But it is hoped that the intrinsic nature of the evolution equations, the lack of domain spatial operators encountered in the Zakharov framework  and, most importantly, the Poisson bracket structure presented could be useful in such investigations. The relation to the KdV bracket and generalizations, mentioned previously, and possible analogies involving wave motions are other directions of interest.

\appendix
\section*{Appendix: Mathematical background.}
Due to the mathematical setting of the problem, a minimal amount of familiarity with Poisson brackets and variational calculus on infinite-dimensional manifolds is assumed.  A brief background is  provided below, but for more background the reader may refer to \cite{MaRa1999, La1970}. It is useful to keep in mind that all ideas are essentially generalizations of ideas applied to finite-dimensional manifolds.\\

  The phase space of the vortex sheet dynamical system is an infinite-dimensional manifold. For the purpose of this paper, it suffices to think of this space as an abstract mathematical surface with `coordinates' being the dynamic variables. As the sheet evolves in the physical domain from a set of initial conditions, it traces out a curve on $P$ starting from an initial point $p \in P$ corresponding to the chosen initial conditions.  At each point of the curve, there is also a tangent vector representing, in a general sense, the `velocity' with which the system is evolving. The collection of all such tangent vectors, associated with all system curves, is termed the vector field $X$ of the dynamical system on $P$. In the case that the dynamical system is a Hamiltonian system, it is termed the Hamiltonian vector field $X_H$ of the system. 

A real-valued function $F$ on $P$ is denoted by $F : P \rightarrow \mathbb{R}$. A functional derivative of $F$ at a point $p \in 
P$ generalizes the notion of a partial derivative of a function on a finite-dimensional space. 
Recall the close relation between the differential of a function and its partial derivatives. Taking as a simple example a function on the $x$-$y$ plane, $f : \mathbb{R}^2 \rightarrow \mathbb{R}$,
\begin{align*}
df (x, y)&= ( \frac{\partial f}{\partial x} dx + \frac{\partial f}{\partial y}dy), \\
&= \left<\nabla f,d\mathbf{r}\right>_{\mathbb{R}^2}
\end{align*}
where $(dx, dy)$ are the components of any infinitesimal vector dr based at point $(x, y)$ and tan- 
gent to the plane. The rate of change of $f$ following a parametrized curve $d(t) \equiv (x(t), y(t))$, 
representing the trajectory of a dynamical system in $\mathbb{R}^2$, is given by
\begin{align*}
\frac{df}{dt}(x, y) = \left(\frac{\partial f}{\partial x}\frac{dx}{dt}+\frac{\partial f}{\partial y}\frac{dy}{dt} \right)_{(x,y)}
\end{align*}
with $(dx/dt, dy/dt)$ the components of the velocity vector, tangent to the curve, at point $(x, y)$. 
The standard Euclidean inner product in the plane $\left \langle \; ,\;\right\rangle_{\mathbb{R}^2}$ is used in the above relations in pairing 
the partial derivatives and the tangent vector.
On an infinite dimensional manifold P the above framework generalizes as follows. Instead of $(x, 
y)$ coordinates, one now has $(X, Y )$ coordinates where the pair $(X, Y )_{\mid p}$ could represent vector 
fields or a real-valued functions based on $C$, and will be denoted by ($X_p, Y_p)$, respectively. A 
real-valued function $F : P \rightarrow R$ is expressed as
\begin{align*}
F (p)&=\oint_C f (X_p(s), Y_p(s)) ds,  \quad f : C \rightarrow \mathbb{R}.
\end{align*}
The inner product pairing in $\mathbb{R}^2$ generalizes to an integral pairing, where the domain of inte- 
gration is $C$. The expression for the differential of F is now given by
\begin{align}
dF (X, Y )&=\oint_C \left( \left<\frac{ \delta f }{\delta X_p}  (s), \delta X (s) \right> + \left<\frac{ \delta f }{\delta Y_p}  (s), \delta Y (s) \right>\right)\;ds, \label{eq:diffF}
\end{align}				
where the pair $(\delta X_p, \delta Y_p)$ is essentially the same type of mathematical object on $C$ as $(X, Y )_{\mid p}$, 
respectively, and is associated with the vector $(\delta X, \delta Y )$ that lies in the (infinite-dimensional) 
tangent plane to $P$ at $p$. The pairing $\left\langle \; , \; \right \rangle $ is the standard Euclidean inner product for vector 
fields and scalar multiplication for real-valued functions.
The expression for the rate of change of $F$ along a curve $d(t) \equiv (X(t), Y (t))$, representing the 
trajectory of a dynamical system in $P$, is then given by
\begin{align*}
\frac{dF}{dt}(X, Y )&= \oint_C \left(\left \langle \frac{\delta f}{\delta X_p}  (s),  \frac{dX_p(s, t) }{dt }\right \rangle+\left \langle \frac{\delta f}{\delta Y_p}  (s),  \frac{dY_p(s, t) }{dt }\right \rangle \right) ds,
\end{align*}
					
Consider now the space of all real-valued functions on P , which is also an infinite-dimensional 
space. Denote it by $\mathcal{F}(P)$. A Poisson bracket is a bilinear operator that acts on pairs of real- 
valued functions to return another real-valued function. Mathematically,$ \left\{ , \right\} : \mathcal{F}(P ) \times  \mathcal{F}(P ) \rightarrow  \mathcal{F}(P )$. Formally, it satisfies four properties: bilinearity, skew-symmetry, Leibniz’s rule and Jacobi’s identity. With the choice of a Poisson bracket, $P$ is termed a Poisson manifold. 
A standard and simple example is the following `canonical' Poisson bracket on $\mathbb{R}^2$:
\begin{align*}
\left\{f , g \right\}(x, y)&=\frac{\partial f (x, y)}{\partial x} \frac{\partial g (x, y)}{\partial y} -\frac{\partial g (x, y)}{\partial x} \frac{\partial f (x, y)}{\partial y}=: m(x, y)
\end{align*}
The bracket thus acts on functions $f$ and $g$, and returns function $m$. 

On infinite-dimensional $P$, this bracket generalizes as: 
\begin{align}
\left\{F, G \right\}(X, Y)&=\oint_C\left( \left<\frac{ \delta f }{\delta X_p}  (s), \frac{\delta g}{\delta Y}(s) \right> - \left<\frac{ \delta g }{\delta X_p}  (s), \frac{\delta f}{\delta Y} (s) \right>\right)\;ds=: h(X, Y) \label{eq:canpb}
\end{align}
Given a Hamiltonian function $H: P \rightarrow \mathbb{R}$ of the form $H(p)=\oint_C h (X_p(s), Y_p(s)) ds$ one then obtains the Hamiltonian vector field $X_H$ relative to the above Poisson bracket, which gives the differential equations of the Hamiltonian system, as follows: for any $F: P \rightarrow \mathbb{R}$,
\begin{align}
&\frac{dF}{dt}= \left\{F,H\right\} \label{eq:pbgen}\\
\Rightarrow & \oint_C \left(\left \langle \frac{\delta f}{\delta X_p}  (s),  \frac{dX_p(s, t) }{dt }\right \rangle+\left \langle \frac{\delta f}{\delta Y_p}  (s),  \frac{dY_p(s, t) }{dt }\right \rangle \right) ds \nonumber \\
& \hspace{1in}=\oint_C\left( \left<\frac{ \delta f }{\delta X_p}  (s), \frac{\delta h}{\delta Y_p}(s) \right> - \left<\frac{ \delta h }{\delta X_p}  (s), \frac{\delta f}{\delta Y_p} (s) \right>\right)\;ds  \nonumber
\end{align}
leading to the equations of motion of the system,
\begin{gather} \label{eq:eomgen}
\begin{aligned}
\frac{dX_p }{dt }&=\frac{\delta h}{\delta Y_p} \\
\frac{dY_p }{dt }&=-\frac{\delta h}{\delta X_p} 
\end{aligned}
\end{gather}
\paragraph{Remark.} The general formula  (\ref{eq:pbgen}) for obtaining the Hamiltonian vector field is the same for any choice of Poisson bracket and $H$. Note that, in particular, with $F=H$, one obtains the well-known fact that $H$ is conserved along system trajectories. Zakharov used the Poisson bracket in (\ref{eq:canpb}), the vortex sheet bracket in this paper is slightly different. Obviously, a different choice of Poisson bracket will  lead to equations of motion that have a different form than (\ref{eq:eomgen}).

\paragraph{Acknowledgements.} The author is grateful to Rangachari Kidambi for many illuminating discussions on the topic of vortex sheets.

\end{document}